\documentclass[floats,prd,aps,showpacs,twocolumn,amssymb,nofootinbib]{revtex4}
\usepackage{amssymb}
\usepackage{graphics}
\usepackage{amsmath}
\usepackage{amsfonts}
\usepackage{bm}% bold math
\usepackage[]{latexsym}
\usepackage{epsfig}

\newcommand{\be}{\begin{equation}}\newcommand{\ee}{\end{equation}}
\newcommand{\bea}{\begin{eqnarray}}\newcommand{\eea}{\end{eqnarray}}
\newcommand{\brr}{\begin{array}}\newcommand{\err}{\end{array}}
\newcommand{\bit}{\begin{itemize}}\newcommand{\eit}{\end{itemize}}
\newcommand{\ben}{\begin{enumerate}}\newcommand{\een}{\end{enumerate}}

\newcommand{\ba}{\begin{array}}
\newcommand{\ea}{\end{array}}

\def\al{\alpha}

\def\1{{_{1}}}\def\2{{_{2}}}

\def\noHe0{:\;\!\!\;\!\!:H_e(0):\;\!\!\;\!\!:}
\def\noHm0{:\;\!\!\;\!\!:H_\mu(0):\;\!\!\;\!\!:}
\def\al{\alpha}

\def\1{{_{1}}}\def\2{{_{2}}}

\begin{document}
\title{Casimir effect in Post-Newtonian Gravity with Lorentz-violation}

\author{M. Blasone\footnote{	blasone@sa.infn.it }$^{\hspace{0.3mm}1,2}$, G. Lambiase\footnote{lambiase@sa.infn.it}$^{\hspace{0.3mm}1,2}$, L. Petruzziello\footnote{lpetruzziello@na.infn.it}$^{\hspace{0.3mm}1,2}$ and An. Stabile\footnote{astabile@unisa.it}$^{\hspace{0.3mm}1,2}$} \affiliation
{$^1$Dipartimento di Fisica, Universit\`a di Salerno, Via Giovanni Paolo II, 132 I-84084 Fisciano (SA), Italy.\\ $^2$INFN, Sezione di Napoli, Gruppo collegato di Salerno, I-84084 Fisciano (SA), Italy.}

\date{\today}
  \def\be{\begin{equation}}
\def\ee{\end{equation}}
\def\al{\alpha}
\def\bea{\begin{eqnarray}}
\def\eea{\end{eqnarray}}

\begin{abstract}

We study the Casimir effect in the framework of Standard Model Extension (SME). Employing the weak field approximation, the vacuum energy density $\varepsilon$ and the pressure for a massless scalar field confined between two nearby parallel plates in a static spacetime background are calculated. In particular, through the analysis of $\varepsilon$, we speculate a constraint on the Lorentz-violating term  $\bar{s}^{00}$ which is lower than the bounds currently available for this quantity. After that, the correction to the pressure given by the gravitational sector of SME is presented. Finally, we remark that our outcome has an intrinsic validity that goes beyond the treated case of a point-like source of gravity.

\end{abstract}
 \vskip -1.0 truecm
\maketitle

\section{Introduction}
\setcounter{equation}{0}

It is well known that General Relativity (GR) exhibits serious incompatibilities when it comes to link its domain of validity with the realm of Quantum Field Theory. The hope is to overcome these obstacles, so that it would be possible to describe the behavior of any interaction including gravity even when quantum effects
are not negligible. In this direction, many reasonable and solid proposals have been made (such as Loop Quantum Gravity and String Theory), but the sensation
is that there is still a great amount of conceptual problems and obstacles to overcome. However, there is a shared and accepted awareness that
allows us to look for a unified theory at Planck scale (namely $m_P\simeq 10^{19}$ GeV); this fact may not be surprising from a theoretical point of view anymore, but practically it means that it is impossible to detect even the smallest signal of quantum gravity. In other words, experiments do not provide any criterion to discern whether a physical argument can be rejected or not at those energy levels.

Nevertheless, even in current laboratory tests, there is still the possibility to search for little traces that can be directly related to an underlying unified theory, and one of the most important concepts in this perspective is represented by Lorentz symmetry breaking. Such a violation is highly recurrent in many candidates of quantum gravity, and for this reason it is considered an essential notion to take into account for a natural extension of our knowledge in such an unknown domain. Actually, Lorentz violation has widely been used as one of the main bedrocks on which to develop physics beyond the Standard Model (SM). SME (Standard Model Extension~\cite{Kost1,Kost4,newkost,Kost2}) is therefore born within this environment, and it is considered one of the most important effective field theories that includes SM as a limiting case. The intuition at the basis of SME comes from the study of covariant string field theory~\cite{string}. The idea is to build all possible scalars of the SME Lagrangian by contracting SM and gravitational fields with suitable coefficients that induce Lorentz (and CPT) violation. Of course, we expect these coefficients to be heavily suppressed, and thus to be considered extremely small if analyzed at current scales. However, many focused experiments have been performed to put constraints on their values and to gather useful information on them~\cite{Kost5}. For an accurate overview on SME, see Ref.~\cite{bluhm}.

In this paper, we consider the Casimir effect in curved spacetime, where the metric is deduced by the gravitational sector of the SME Lagrangian. Generally speaking, the Casimir effect~\cite{casimir1,casimir2} arises when a quantum field is bounded in a finite space. Such a confinement reduces the modes of the quantum field producing, as a consequence, a measurable manifestation. The Casimir effect has been studied in flat spacetime in great detail~\cite{casimir_exp1,nesterenkoC,nesterenkoC2,nesterenkoC3,nesterenkoC4,nesterenkoC5,nesterenkoC6,nesterenkoC7,casimir_exp2,casimir_exp3, casimir_exp4}, showing the robustness of the assumptions at the basis of its theoretical explanation. In this framework, there are already works that study the Casimir effect with the contribution of the SME coefficients for the fermion and photon sector~\cite{frank,escobar}. In some recent papers~\cite{setare, calloni, caldwell,sorge,EspNapRos, BimEspRos, Calloni14, Borzoo, BuoCan,TanPir,lamb}, instead, the analysis of the role of a gravitational field in the vacuum energy density of a quantum field inside a cavity has been performed. This opened the doors to many interesting developments and lines of research. Indeed, an interesting investigation on the consistency between the Casimir energy and the equivalence principle is conduced in Refs.~\cite{Fulling}. Moreover, possible modifications in the vacuum energy could become relevant in the dynamics of the universe~\cite{caldwell,brevik}. Microscopically, modifications of Casimir's energy could be crucial in the context of quark confinement based on string interquark potentials~\cite{lambiase,lambiase2,lambiase3,lambiase4}. Finally, the implications of gravity on Casimir effect deal with the open issue regarding the limits of validity of GR at small distances~\cite{mostepanenko}.

The aims of this article are essentially two:
\begin{itemize}
\item[--] to obtain a significant and plausible constraint on SME Lorentz-violating terms derived within a Post-Newtonian expansion
of the metric tensor describing the spacetime in proximity of a point-like source of gravity \cite{Kost3};
\item[--] to see how the pressure between the Casimir plates changes in the above context.
\end{itemize}
In order to do this, we analyze the dynamics of a massless scalar field in the context of Casimir effect, employing a technique already used in the search for direct evidences of extended theories of gravity \cite{lamb,FOG_CGL2,stabile_stabile_cap,stabstab,FOG_CGL}.
Klein-Gordon equation between the two Casimir plates shall be solved, assuming that the distance between them is much smaller than the distance between the source of the gravitational field and the plates.

The paper is organized as follows: in Sec.~II the metric tensor is presented, as derived in the Post-Newtonian approximation of the purely gravitational sector of SME in Ref.~\cite{Kost3}. In Sec.~III the dynamics of a canonic massless scalar field is studied within the Casimir plates, and in Sec.~IV the bound and the expression for the pressure are obtained. Discussions and conclusions are given in Sec.~V.

\section{Metric tensor for a point-like source with Lorentz-violating terms}
\label{sect1}
The most general Lagrangian density for the SME gravitational sector of Ref.~\cite{Kost2} contains both a Lorentz-invariant and a Lorentz-violating term.
%\be\label{lagrangian}
%\mathcal{L}_{grav}=\mathcal{L}_{LI}+\mathcal{L}_{LV}.
%\ee
The background is represented by a Riemann-Cartan spacetime, but for our purposes we take the limit of vanishing torsion, in such a way that the Lorentz-invariant part is the usual Einstein-Hilbert contribution. The effective action in which we consider only the leading-order Lorentz-violating
terms is thus given by
\be
S=S_{EH}+S_{LV}+S_{m},
\ee
where
\begin{equation}
S_{EH}=\frac{1}{2\kappa}\int d^{4}x\,\sqrt{-g}\,\mathcal{R},
\end{equation}
is the aforementioned Einstein-Hilbert action, with $\kappa=8\pi G$, $S_{m}$ the matter action and $S_{LV}$ the Lorentz-violating term:
\begin{equation}\label{lfields}
S_{LV}=\frac{1}{2\kappa}\int d^{4}x\,\sqrt{-g}\left(-u\,\mathcal{R}+s^{\mu\nu}\,R_{\mu\nu}^{T}+t^{\rho\lambda\mu\nu}\,C_{\rho\lambda\mu\nu}\right).
\end{equation}
Here, $\mathcal{R}$ is the Ricci scalar, $R_{\mu\nu}^{T}$ the trace-free
Ricci tensor, $C_{\rho\lambda\mu\nu}$ the Weyl conformal tensor and
all other terms contain the information of Lorentz violation. Of course,
they must depend on spacetime position and have to be treated as
dynamical fields, in order to be compatible with lack of prior geometry, a typical feature of GR\footnote{See Ref.~\cite{mattingly} for a detailed explanation of this concept.}.

Since the fields $u$, $s^{\mu\nu}$ and $t^{\rho\lambda\mu\nu}$ in Eq.~(\ref{lfields}) are the ones responsible for Lorentz violation, they acquire a vacuum expectation value, so that it is possible to write fluctuations around them as
\begin{equation}\label{assumption1}
u=\bar{u}+\tilde{u}, \qquad s^{\mu\nu}=\bar{s}^{\mu\nu}+\tilde{s}^{\mu\nu}, \qquad t^{\rho\lambda\mu\nu}=\bar{t}^{\rho\lambda\mu\nu}+\tilde{t}^{\rho\lambda\mu\nu}.
\end{equation}
Furthermore, following Ref.~\cite{Kost3}, we require that each first element of the r.h.s. of Eq.~(\ref{assumption1}) is constant in asymptotically inertial Cartesian coordinates.
However, the fundamental assumption is that, when dealing with Lorentz violation, one always takes into account only the vacuum expectation
values of Eq.~(\ref{assumption1}), completely neglecting fluctuations. This ansatz is reasonable, because we expect to have extremely small deviations from Lorentz symmetry realized in nature.

Without entering the details of calculation given in Ref.~\cite{Kost3}, it is possible to derive the most general metric tensor for a point-like
source of gravity, whose non-null components are
given by
\begin{eqnarray}\nonumber
&&g_{00}=1-\frac{GM}{r}\left(2+3\,\bar{s}^{00}\right),\\[2mm]
&&g_{ij}=\left[-1-\frac{GM}{r}\left(2-\bar{s}^{00}\right)\right]\delta_{ij}.
\label{Metric1}
\end{eqnarray}

\section{Dynamics of a massless scalar field}
\label{sect2}

Let us now consider a conventional massless scalar field $\psi\left(\mathbf{x},t\right)$ in  curved background, i. e. we consider the SME parameters are only into gravity sector. In general, the  $\bar{s}^{\mu\nu}$ parameters can be moved from the gravity sector into the scalar sector using a coordinate choice \cite{newkost}. The choice does not change the physics, so although the calculation looks different it must give the same result.

In our analysis, the Klein-Gordon equation reads~\cite{BirDav}
\begin{equation}\label{FiedlEquation}
\left(\Box+\zeta\,\mathcal{R}\right)\psi\left(\mathbf{x},t\right)=0,
\end{equation}
where $\Box$ is the d'Alembert operator in curved space and $\zeta$
is the coupling parameter between geometry and matter.

\begin{figure}[h]
%\centering
\includegraphics[scale=.5]{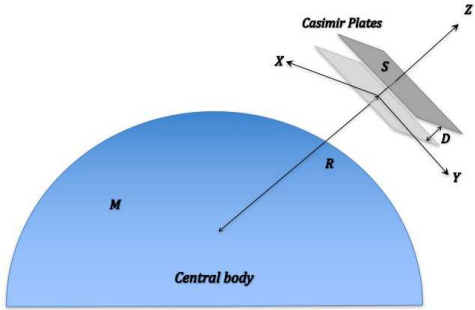}
\caption{The Casimir-like system in a gravitational field is represented above. Here, $D$ denotes the distance between the plates, $S$ their surface and $R$ the distance from the source of gravity of mass $M$, with $D<\sqrt{S}\ll R$.}
\label{fig_1}
\end{figure}

As it can be seen in Fig.~(\ref{fig_1}), the configuration is simple: the
plates are set in such a way that the one nearer to
the source of gravity is distant $R$ from it,
and hence we can choose Cartesian coordinates so that $r=R+z$, where
the variable $z$ is free to vary in the interval $\left[0,\,D\right]$,
if we denote with $D$ the separation between the plates. Clearly,
the relation $D\ll R$ holds.

A further simplification comes from the fact that the only Cartesian coordinate
explicitly present in the quantities appearing in Eq.~(\ref{FiedlEquation}) is $z$.
In fact, denoting $\phi=-\frac{GM}{R}$ and recalling that $\frac{z}{R}\ll1$, the metric tensor becomes
\begin{eqnarray}\nonumber
g_{00}&\approx & 1-\phi\left(1-\frac{z}{R}\right)\left(2+3\,\bar{s}^{00}\right),\\[2mm]
g_{ij}&\approx &\left[-1-\phi\left(1-\frac{z}{R}\right)\left(2-\bar{s}^{00}\right)\right]\delta_{ij}\,,
\end{eqnarray}
with the scalar curvature that assumes the form\footnote{The presence of Lorentz-violating terms allows for a non-vanishing scalar curvature. However, details of its form will not be necessary in the next steps, since it contributes to the mean vacuum energy density only at higher orders.} $\mathcal{R}\equiv\mathcal{R}_1+z\,\mathcal{R}_2$.

At this point, it is clear that the interest is focused on the variation
of the field along the radial direction (namely, along the z-axis). Because there is no explicit dependence on other coordinates, one can think of a solution of the form $\psi\left(\mathbf{x},t\right)=Ne^{i\left(\omega t-\mathbf{k}_{\perp}\cdotp\mathbf{x}_{\perp}\right)}\varphi\left(z\right)$,
where $\mathbf{k}_{\perp}=\left(k_{x},k_{y}\right)$, $\mathbf{x}_{\perp}=\left(x,y\right)$
and $N$ is the normalization factor.

The field equation can thus be rewritten as
\begin{equation}
\partial_{z}^{2}\,\varphi+C_1\,\partial_{z}\,\varphi+C_2\,\varphi=0\,,
\end{equation}
where
\begin{equation}
C_1=-2\,\frac{\phi}{R}\,\bar{s}^{00}, \qquad C_2=a+b\, z\,,
\end{equation}
with
\be\nonumber
a=\omega^{2}\left[1-2\,\phi\left(\bar{s}^{00}+2\right)\right]+\zeta\,\frac{\phi}{R^{2}}\left(4-10\,\bar{s}^{00}\right)-\mid\mathbf{k}_{\perp}\mid^{2},
\ee
\vspace{-4.5mm}\be
b=4\,\frac{\phi}{R}\left(\omega^{2}-3\,\frac{\zeta}{R^{2}}\right).
\ee
The solution of this differential equation is a linear
combination of Airy functions of the first and of the second kind,
with argument $x\left(z\right)=\left[\frac{1}{4}\left(C_1^{2}-4\,a\right)-b\, z\right]\left(-b\right)^{-\frac{2}{3}}.$
In the considered approximation, the solution can be written as
\begin{equation}
\varphi\left(z\right)= k_{1}\,\mathrm{Ai}\left(\frac{-a-b\, z}{\left(-b\right)^{\frac{2}{3}}}\right)+k_{2}\,\mathrm{Bi}\left(\frac{-a-b\, z}{\left(-b\right)^{\frac{2}{3}}}\right).
\end{equation}
Airy functions can be expressed in terms of Bessel functions~\cite{Zuber}.
Due to the form of $a$ and $b$, it is clear that the argument of the Bessel functions
$\eta(z)\equiv\left[a+ b z\right]\left(-b\right)^{-\frac{2}{3}}\gg1$, and hence their asymptotic behavior
yields
\begin{equation}
\varphi\left(z\right)\approx\sqrt{\frac{3}{\pi\sqrt{\eta(z)}}}\,\sin\left[\frac{2}{3}\eta^{\frac{3}{2}}(z)+\tau\right]\,.
\end{equation}
If we impose the Dirichlet boundary conditions on the plates for the field $\varphi(z)$, that is $\varphi(0)=\varphi(D)=0$, we get the relation $\frac{2}{3} \bigl[\eta^{3/2}(0)-\eta^{3/2}(L)  \bigr]=n\,\pi$, where $n$ is an integer. From these boundary conditions, we find the energy spectrum
\begin{eqnarray}\label{Energy_Spectrum}\nonumber
\omega^2_n\,&=&\,\bigl[1-2\,\phi\,(\bar{s}^{00}+2)+4\,\frac{\phi}{R}\,D\bigr]\biggl[\textbf{k}_\bot^2+\\
&&+\biggl( \frac{n\pi}{D}\biggr)^2\biggr]+
\frac{\zeta\,\phi}{R}\,\biggl[10\,\bar{s}^{00}-4+\,\frac{6\,D}{R}\biggr]\,.
\end{eqnarray}
Finally, using the scalar product defined for quantum fields in curved spacetimes \cite{BirDav},
one derives the normalization constant
\begin{eqnarray}\label{N^2}
N_n^2\,=\,\frac{a}{3\,S\,\,b^{1/3}\omega_n\,n\bigl[1-\phi\left(1+\frac{3}{2}\bar{s}^{00}\right)\bigl]},
\end{eqnarray}
with $S$ being the surface of the plates.

\section{Derivation of the bound and of the pressure}\label{sect3}

In order to calculate the mean vacuum energy density $\varepsilon$ between the plates, we use the general relation \cite{BirDav}
\begin{eqnarray}
\varepsilon\,=\,\frac{1}{V_p}\sum_n\int d^2\,\textbf{k}_\bot\int dx\,dy\,dz\,\sqrt{-g} \,\bigl( g_{00}\bigr)^{-1}\,T_{00},
\end{eqnarray}
where $T_{00}\equiv T_{00}\bigl( \psi_n,\psi_n^*\bigr)$ is a component of the energy-momentum tensor $T_{\mu\nu}\,=\,\partial_{\mu}\,\psi\,\partial_{\nu}\,\psi-\frac{1}{2}g_{\mu\nu}g^{\alpha\beta}\partial_{\alpha}\,\psi\,\partial_{\beta}\,\psi$ and $V_p=\int dx\,dy\,dz\,\sqrt{-g}$ is the proper volume. Using the Schwinger proper-time representation and $\zeta$-function regularization, we find the mean vacuum energy density
\begin{eqnarray}\label{DEV}
&\varepsilon&\,=\,\varepsilon_0+\varepsilon_{GR}+\varepsilon_{LV}\,,\\[2mm]
&\varepsilon_0&\,=\,-\frac{\pi^{2}}{1440\,D_p^{4}},\\[2mm]
\label{DEVGR}
&\varepsilon_{GR}&\,= -\frac{\phi \,D_p}{R}\,\varepsilon_0\,,\\[2mm]
\label{DEVLV}
&\varepsilon_{LV}&\,= -6\,\phi\, \bar{s}^{00}\,\varepsilon_0\,,
\end{eqnarray}
where $\varepsilon_0$ is the standard term of Casimir effect, $\varepsilon_{GR}$ is the contribution due to GR and $\varepsilon_{LV}$ is the Lorentz-violating term, with $D_p=\int dz\,\sqrt{-g}$ being the proper length of the cavity. Note that, in our analysis, we have neglected higher-order contributions.

Equation~(\ref{DEV}) gives us the expression of Casimir vacuum energy density at the second order $\mathcal{O}(R^{-2})$ in the framework of SME.
We note that the part related to GR (Eq.~(\ref{DEVGR})) does not have contributions at the first order in $\mathcal{O}(R^{-1})$, but only at higher orders, such as $\mathcal{O}(R^{-2})$. The Lorentz-violating sector Eq.~(\ref{DEVLV}), instead, exhibits a first order factor in $\mathcal{O}(R^{-1})$ connected to $\bar{s}^{00}$.

To obtain a plausible bound on $\bar{s}^{00}$,  we make the assumption $|\varepsilon_{LV}|\lesssim | \varepsilon_{GR}|$. This agrees with several considerations and results expressed in Refs.~\cite{Kost5, frank, Kost3} and ensures the fact that  Lorentz-violating manifestations are small, as widely employed in Lorentz-violation phenomenology~\cite{Kost3}. However, the reasonableness of the constraint we derive cannot be directly tested, since $\varepsilon$ is still now an unmeasurable quantity. This is why we need to compare the heuristic constraint with a physical one, which can only be calculated using the pressure.

Apart from the previous comment, considering the case of the Earth and requiring $D_p\thicksim 10^{-7}$\,m (a typical choice for the proper length in standard literature) for the plausible assumption exhibited above, we get
\begin{eqnarray}\label{CONSTRAINTENR}
 \bar{s}^{00} \lesssim \frac{D_p}{6\,R_\oplus} \lesssim 10^{-14},
\end{eqnarray}
where $R_\oplus\thicksim 6.4\times 10^{6}$\,m.

It must be pointed out that recent developments in nanotechnology can further strengthen the above bound by one or two orders of magnitude. In fact, in the near future, the value of $D_p$ could reach scales even smaller than nanometers (as already contemplated, for example, in Ref.~\cite{exp}), thus transforming Eq.~(\ref{CONSTRAINTENR}) into a more stringent constraint, $\bar{s}^{00}\lesssim 10^{-15}$.

Let us now turn the attention to the pressure. The attractive force observed between the cavity plates is obtained by the relation $F=-\frac{\partial {\cal E}}{\partial D_p}$, where ${\cal E}=\varepsilon\,V_P$ is the Casimir vacuum energy. Then, the pressure is simply given by $P=F/S$, and hence
\begin{eqnarray}\label{pressure}
P&=&P_{0}+P_{GR}+P_{LV}\,,\\[2mm]
P_{0}&=&-\frac{\pi^{2}}{480D_p^{4}}\,,\label{po}\\[2mm]
P_{GR}&=&3\,\phi\, P_{0}\,,\label{pgr}\\[2mm]
P_{LV}&=&-6\,\phi\,\bar{s}^{00}\,P_{0}\,,\label{plv}
\end{eqnarray}
where $P_{0}$ is the pressure in the flat case, while $P_{GR}$ is the pressure in GR and $P_{LV}$ is the  contribution connected to Lorentz-violation.

We now want to test the compatibility of SME with the experimental data to check how much a concrete bound differs from the heuristic one obtained in Eq.~\eqref{CONSTRAINTENR}. This can be achieved by using the pressure as a measurable physical quantity. In fact, imposing the constraint $|P_{LV}|\lesssim \delta P$, where $\delta P$ is the experimental error, we obtain the following relation:
\begin{equation}\label{Constraint}
\bar{s}^{00}\,\lesssim\frac{\delta P}{P_{0}}\frac{1}{6\phi}=\frac{1}{3}\frac{\delta P}{P_{0}}\frac{R}{R_S}\,,
\end{equation}
where $R_S$ is Schwarzschild radius.

The total absolute experimental error of the measured Casimir pressure~\cite{mostepanenko2} is 0.2\% ($ \delta P/P_0\simeq 0.002$). Typical values of the ratio $\frac{R}{R_S}$ in the Solar System are included between $10^7\div10^{10}$. In particular, for the Earth we have $7.2\times10^{8}$, which means that the term on the r.h.s. of Eq.~(\ref{Constraint}) is of order $10^{6}$. The comparison of such a result with Eq.~\eqref{CONSTRAINTENR} clearly shows that we cannot use\footnote{Unless we believe the heuristic bound to be true and thus physically consistent.} the Casimir experiment to measure the pressure in order to significantly constrain the parameter $\bar{s}^{00}$. To to this, we need to enhance the experimental sensitivity on Earth by at least six order of magnitude, in such a way that $\frac{\delta P}{P}\lesssim 10^{-9}$. Nevertheless, in the near future, gravitational interferometers may provide a valid framework to test SME. Indeed, they have reached a high sensitivity, and therefore they could be, in principle, used as the tool to detect the small effects induced by the the Lorentz-violating contributions on Casimir pressure.

Finally, a word must be spent on the choice of point-like source of gravity. Although it has been considered only to simplify the treatment of the problem, the relevance of the outcome does not depend on it. In fact, even if we considered a rotating spherical object instead of a point, the value of the constraint would basically be the same. Lorentz-violating factors will very likely appear as combinations of the $\bar{s}^{\mu\nu}$ already introduced in this work, but the line of reasoning would exactly be the same. Consequently, we expect Eqs.~(\ref{CONSTRAINTENR}) and \eqref{Constraint} to be modified only in its l.h.s., for example with contributions such as $\bar{s}^{00}+\sum_{\mu,j}\bar{s}^{\mu j}$ at the lowest order.

\section{Conclusions}

In the context of SME, working in the weak field approximation, we have studied the dynamics of a massless scalar field confined between two nearby parallel plates in a static spacetime background generated by a point-like source. In order to obtain a reasonable constraint on Lorentz-violating terms in the context of the Casimir effect, we have derived the corrections to the flat spacetime Casimir vacuum energy density Eq.~(\ref{DEV}), in the framework of SME. We have found that, in the energy density, GR gives us only contributions at the second order $\mathcal{O}(R^{-2})$, while Lorentz-violating corrections occur at first order $\mathcal{O}(R^{-1})$. After that, we have evaluated the pressure Eq.~\eqref{pressure} to observe how it changes from the usual expression in flat spacetime, Eq.~\eqref{po}, in the presence of gravity (see Eq.~\eqref{pgr}) and with SME coefficients (see Eq.~\eqref{plv}).

By requiring $|\varepsilon_{LV}|\lesssim | \varepsilon_{GR}|$, we have then been able to find a significant bound on the SME coefficient $\bar{s}^{00}$. Such an assumption is related to the fact that manifestations of Lorentz violation in nature are expected to be extremely evanescent. If the above inequality did not hold true, it would have been possible to detect traces of Lorentz-violating terms in the tests proposed in Ref.~\cite{Kost3} and in other experiments involving the intertwining between SME and gravity, but this is not the case.

We remark that, for the problem at hand, there is the necessity to have a direct access to the vacuum energy density in order to evaluate $\bar{s}^{00}$. Actually, as already pointed out, the true measurable physical quantity in the context of the Casimir effect is the pressure $P$. One can possibly extract a constraint for $\bar{s}^{00}$ also with $P$ as done in Eq.~\eqref{Constraint}, but its order of magnitude would be extremely high if compared to Eq.~(\ref{CONSTRAINTENR}) and especially to the data of Ref.~\cite{Kost5}. The current technology is far from allowing a direct experimental check of the influence of SME on Casimir effect. Nevertheless, in the near future, gravitational interferometers might achieve high sensitivity, providing an alternative tool for testing the SME with a more stringent and efficient bounds reachable through the evaluation of the pressure.

We also point out that, following the same analysis of Refs.~\cite{Fulling}, no violation of equivalence principle arises in our framework, i.e. the parameter space here analyzed leads to the conclusion that the coefficients $\bar{s}^{\mu\nu}$ do not allow to discriminate between inertial mass and gravitational mass.

Finally, the consideration after Eq.~(\ref{CONSTRAINTENR}) is corroborated by the fact that, assuming we have to deal with the Kerr metric for a more detailed analysis, the off-diagonal contribution related to the angular momentum $J$, that is $\frac{G J_\oplus}{R_\oplus c^3}$, has the same order of magnitude of the diagonal term
\be
\label{esteem}
\frac{2 G M_\oplus}{R_\oplus c^2}\thicksim 10^{-9}\,.
\ee
This knowledge tells us that, in Eq.~(\ref{DEVGR}), we would obtain an analytically different expression, $\varepsilon_{GR}=f(M, D_p, R, J)\,\varepsilon_0$, which nonetheless should possess the same order of magnitude of Eq.~(\ref{CONSTRAINTENR}). Moveover, outcomes of a recent work \cite{sorge2} suggest that, in the case of a geostationary orbit, the effects of rotation  on the mean vacuum energy density can be completely neglected, due to the fact that the Casimir-like system acquires the same angular velocity of the Earth. These hints strengthen the conjecture that implications of the current work are basically untouched also in a more realistic treatment of the studied phenomenon, which however will be the object of future investigation.

\medskip
\section*{Acknowledgments}
The authors would like to thank V. A. Kosteleck\'y and E. Calloni for useful discussions and comments.


\begin{thebibliography}{99}

\bibitem{Kost1} D. Colladay and V. A. Kosteleck\'y, Phys. Rev. D \textbf{55}, 6760 (1997).
%CPT violation and the Standard Model

\bibitem{Kost4} D. Colladay and V. A. Kosteleck\'y, Phys. Rev. D \textbf{58}, 116002 (1998).
%Lorentz-violating extension of the Standard Model

\bibitem{newkost} V. A. Kosteleck\'y and J. D. Tasson, Phys. Rev. D \textbf{83}, 016013 (2011).

\bibitem{Kost2} V. A. Kosteleck\'y, Phys. Rev. D \textbf{69}, 105009 (2004).
%Gravity, Lorentz violation and the Standard Model

\bibitem{string} V. A. Kosteleck\'y and S. Samuel, Phys. Rev. D \textbf{39}, 683 (1989).
%Spontaneous breaking of Lorentz symmetry in string theory

\bibitem{Kost5} V. A. Kosteleck\'y and N. Russell, Rev. Mod. Phys. \textbf{83}, 11 (2011).
%Data tables for Lorentz and CPT violation

\bibitem{bluhm} R. Bluhm, Lect. Notes Phys. \textbf{702}, 191-226 (2006).
%Overview of the SME: Implications and Phenomenology of Lorentz violation

\bibitem{casimir1} H. Casimir, Proc. K. Ned. Akad. Wet. {\bf 51} 793 (1948).

\bibitem{casimir2} H. Casimir, PolderD. Phys. Rev. {\bf 73} 360 (1948).

\bibitem{casimir_exp1} K. A. Milton, \emph{The Casimir effect: Physical Manifestations of Zero-Point Energy}, River edge: World Scientific, 2001.

\bibitem{nesterenkoC}
V.V. Nesterenko, G. Lambiase and G. Scarpetta, Riv. Nuovo Cim. {\bf 27}, N6, 1-74 (2004).

\bibitem{nesterenkoC2}
 V.V. Nesterenko, G. Lambiase and G. Scarpetta, Annals Phys. {\bf 298}, 403 (2002).

\bibitem{nesterenkoC3}
        V.V. Nesterenko, G. Lambiase and G. Scarpetta, Int. J. Mod. Phys. A {\bf 17}, 790 (2002).

\bibitem{nesterenkoC4}
        V.V. Nesterenko, G. Lambiase and G. Scarpetta, Phys. Rev. D {\bf 64}, 025013 (2001).

\bibitem{nesterenkoC5}
        V.V. Nesterenko, G. Lambiase and G. Scarpetta, J. Math. Phys. {\bf 42}, 1974 (2001).

\bibitem{nesterenkoC6}
        G. Lambiase, G. Scarpetta and V.V. Nesterenko, Mod. Phys. Lett. A {\bf 16}, 1983 (2001).

\bibitem{nesterenkoC7}
        G. Lambiase, V.V. Nesterenko and M. Bordag, J. Math. Phys. {\bf 40}, 6254 (1999).

\bibitem{casimir_exp2} M. Bordag, U. Mohideen and V. M. Mostepanenko, Phys. Rep. {\bf 353}, 1 (2001).

\bibitem{casimir_exp3} C. Genet, A. Lambrecht and S. Reynaud, \emph{On the Nature of Dark Energy} 18th IAP Coll. on the Nature of Dark Energy: Observations and Theoretical Results in the Accelerating Universe, Paris, France, 1-5 July 2002, ed P. Brax, J. Martin, J. P. Uzan (Fronter Group) pp 121-30.

\bibitem{casimir_exp4} G. Bressi, G. Carugno, R. Onofrio and G. Ruoso, Phys. Rev. Lett. {\bf 88}, 041804 (2002).

\bibitem{frank} M. Frank and I. Turan, Phys. Rev. D \textbf{74}, 033016 (2006).

\bibitem{escobar} A. Martin-Ruiz and C. A. Escobar, Phys. Rev. D \textbf{94}, 076010 (2016).

\bibitem{setare} M. R. Setare, Class. Quantum Grav. {\bf18}, 2097 (2001).

\bibitem{calloni} E. Calloni, L. di Fiore, G. Esposito, L. Milano and L. Rosa, Int. J. Mod. Phys. A {\bf 17}, 804 (2002).

\bibitem{caldwell} R. R. Caldwell, 	arXiv:astro-ph/0209312.

\bibitem{sorge} F. Sorge, Class. Quantum Grav. {\bf 22}, 5109 (2005).

\bibitem{EspNapRos} G. Esposito, G. M. Napolitano and L. Rosa, Phys.\ Rev.\ D {\bf 77}, 105011 (2008).

\bibitem{BimEspRos}  G. Bimonte, G. Esposito and L. Rosa, Phys.\ Rev.\ D {\bf 78}, 024010 (2008).

\bibitem{Calloni14} E. Calloni, M. De Laurentis, R. De Rosa, F. Garufi, L. Rosa, L. Di Fiore, G. Esposito, C. Rovelli, P. Ruggi, and F. Tafuri, Phys.\ Rev.\ D {\bf 90}, 022002 (2014).

\bibitem{Borzoo} B. Nazari,  Eur. Phys. J. C \textbf{75}, 501 (2015).

\bibitem{BuoCan} P. Bueno, P. A. Cano, V. S. Min and M. R. Visser, Phys.\ Rev.\ D {\bf 95}, 044010 (2017).

\bibitem{TanPir} M. R. Tanhayi and R. Pirmoradian, R. Int. J. Theor. Phys. \textbf{55}, 766 (2016).

\bibitem{lamb} G. Lambiase, A. Stabile and An. Stabile, Phys. Rev. D \textbf{95}, 084019 (2017).
%Casimir effect in extended theories of gravity

\bibitem{Fulling}
 S.~A.~Fulling, K.~A.~Milton, P.~Parashar, A.~Romeo, K.~V.~Shajesh and J.~Wagner,
  %``How Does Casimir Energy Fall?,''
  Phys.\ Rev.\ D {\bf 76}, 025004 (2007);
  K.~A.~Milton, P.~Parashar, K.~V.~Shajesh and J.~Wagner,
  %``How does Casimir energy fall? II. Gravitational acceleration of quantum vacuum energy,''
  J.\ Phys.\ A {\bf 40}, 10935 (2007);
   K.~V.~Shajesh, K.~A.~Milton, P.~Parashar and J.~A.~Wagner,
  %``How does Casimir energy fall? III. Inertial forces on vacuum energy,''
  J.\ Phys.\ A {\bf 41}, 164058 (2008);
   K.~A.~Milton, K.~V.~Shajesh, S.~A.~Fulling and P.~Parashar,
  %``How does Casimir energy fall? IV. Gravitational interaction of regularized quantum vacuum energy,''
  Phys.\ Rev.\ D {\bf 89}, 064027 (2014);
    K.~A.~Milton, S.~A.~Fulling, P.~Parashar, A.~Romeo, K.~V.~Shajesh and J.~A.~Wagner,
  %``Gravitational and inertial mass of Casimir energy,''
  J.\ Phys.\ A {\bf 41}, 164052 (2008).

\bibitem{brevik} I. Brevik, K. A. Milton, S. D. Odintsov and K. E. Osetrin, Phys. Rev. {\bf 62}, 064005 (2000).

\bibitem{lambiase} G. Lambiase and V. V. Nesterenko, Phys Rev. D {\bf 54}, 6387 (1996).

\bibitem{lambiase2}
            L. Hadasz, G. Lambiase and V.V. Nesterenko, Phys. Rev. D {\bf 62}, 025011 (2000).

\bibitem{lambiase3}
 	       G. Lambiase and V.V. Nesterenko, Phys. Lett. B {\bf 398}, 335 (1997).
 	
\bibitem{lambiase4}
            H. Kleinert, G. Lambiase and V.V. Nesterenko, Phys. Lett. B {\bf 384}, 213 (1996).

\bibitem{mostepanenko} V. M. Mostepanenko and M. Novello, arXiv:hep-ph/0008035.

\bibitem{Kost3} Q. G. Bailey and V. A. Kosteleck\'y, Phys. Rev. D \textbf{74}, 045001 (2006).
%Signals for Lorentz violation in post-Newtonian gravity

%\bibitem{mostepanenko2} V. M. Mostepanenko, J. Phys. Conf. Ser. {\bf 161}, 012003 (2009).

\bibitem{FOG_CGL2}
  G. Lambiase, M. Sakellariadou, A. Stabile and An. Stabile, JCAP \textbf{1507}, 003 (2015).

  \bibitem{stabile_stabile_cap}
          A. Stabile, An. Stabile and S. Capozziello, Phys. Rev. D {\bf 88}, 124011 (2013).

          \bibitem{stabstab} A. Stabile and An. Stabile, Phys. Rev. D {\bf 85}, 044014 (2012).

          \bibitem{FOG_CGL}
S. Capozziello, G. Lambiase, M. Sakellariadou, A. Stabile and An. Stabile, Phys.Rev. D {\bf 91}, 044012 (2015).

\bibitem{mattingly} D. Mattingly, Living Rev. Rel. \textbf{8}, 5 (2005).
%Modern tests of Lorentz invariance

%\bibitem{riess}
 %        A.G. Riess  {\it et al.},
%Observational Evidence from Supernovae for an Accelerating Universe.
%{\it The Astronomical Journal} 1998, {\bf 116}, 1009-1038.

\bibitem{Zuber}
 I. S. Gradshteyn and I. M. Ryzhik, \emph{Table of Integrals, Series and Products}, New York: Academic, 1980.

\bibitem{BirDav} N. D. Birrell and P. C. W. Davies, \emph{Quantum Fields in Curved Space}, Cambridge: Cambridge University
Press, 1982.

\bibitem{exp} A. Manjavacas, F. J. Rodriguez-Fortuno, F. J. Garcia de Abajo, and A. V. Zayats,
Phys. Rev. Lett. \textbf{118}, 133605 (2017).

\bibitem{mostepanenko2} V. M. Mostepanenko, Experiment, J. Phys. Conf. Ser. \emph{161}, 012003 (2009).

\bibitem{sorge2} F. Sorge,
%Casimir energy in Kerr space-time
Phys. Rev. D {\bf 90}, 084050 (2014).

\end{thebibliography}
\end{document}